\documentclass[%
 reprint,
superscriptaddress,
showpacs,preprintnumbers,
 amsmath,amssymb,
 aps,
]{revtex4-1}

\usepackage{graphicx}
\usepackage{dcolumn}
\usepackage{bm}

\begin{document}


\title{Search for unbound $^{15}$Be states in the $3n+^{12}$Be channel}

\author{A. N. Kuchera}
\email{anthony.kuchera@gmail.com}
\affiliation{National Superconducting Cyclotron Laboratory, Michigan State University, East Lansing, Michigan 48824, USA}

\author{A. Spyrou}
\affiliation{National Superconducting Cyclotron Laboratory, Michigan State University, East Lansing, Michigan 48824, USA}
\affiliation{Department of Physics \& Astronomy, Michigan State University, East Lansing, Michigan 48824, USA}

\author{J. K. Smith}
\altaffiliation[Present Address: ]{TRIUMF, 4004 Wesbrook Mall, Vancouver, British Columbia, V6T 2A3 Canada}
\affiliation{National Superconducting Cyclotron Laboratory, Michigan State University, East Lansing, Michigan 48824, USA}
\affiliation{Department of Physics \& Astronomy, Michigan State University, East Lansing, Michigan 48824, USA}

\author{T. Baumann}
\affiliation{National Superconducting Cyclotron Laboratory, Michigan State University, East Lansing, Michigan 48824, USA}

\author{G. Christian}
\altaffiliation[Present Address: ]{TRIUMF, 4004, Wesbrook Mall, Vancouver, British Columbia, V6T 2A3 Canada}
\affiliation{National Superconducting Cyclotron Laboratory, Michigan State University, East Lansing, Michigan 48824, USA}
\affiliation{Department of Physics \& Astronomy, Michigan State University, East Lansing, Michigan 48824, USA}

\author{P. A. DeYoung}
\affiliation{Department of Physics, Hope College, Holland, Michigan 49422, USA}

\author{J. E. Finck}
\affiliation{Department of Physics, Central Michigan University, Mount Pleasant, Michigan 48859, USA}

\author{N. Frank}
\affiliation{Department of Physics \& Astronomy, Augustana College, Rock Island, Illinois 61201, USA}

\author{M. D. Jones}
\affiliation{National Superconducting Cyclotron Laboratory, Michigan State University, East Lansing, Michigan 48824, USA}
\affiliation{Department of Physics \& Astronomy, Michigan State University, East Lansing, Michigan 48824, USA}

\author{Z. Kohley}
\affiliation{National Superconducting Cyclotron Laboratory, Michigan State University, East Lansing, Michigan 48824, USA}
\affiliation{Department of Chemistry, Michigan State University, East Lansing, Michigan 48824, USA}

\author{S. Mosby}
\altaffiliation[Present Address: ]{LANL, Los Alamos, New Mexico 87545, USA}
\affiliation{National Superconducting Cyclotron Laboratory, Michigan State University, East Lansing, Michigan 48824, USA}
\affiliation{Department of Physics \& Astronomy, Michigan State University, East Lansing, Michigan 48824, USA}

\author{W. A. Peters}
\altaffiliation[Present Address: ]{Department of Physics \& Astronomy, University of Tennessee, Knoxville, Tennessee 37996, USA}
\affiliation{Department of Physics \& Astronomy, Rutgers University, Piscataway, New Jersey 08854, USA}

\author{M. Thoennessen}
\affiliation{National Superconducting Cyclotron Laboratory, Michigan State University, East Lansing, Michigan 48824, USA}
\affiliation{Department of Physics \& Astronomy, Michigan State University, East Lansing, Michigan 48824, USA}

\date{\today}

\begin{abstract}
\begin{description}
\item[Background] $^{15}$Be is expected to have low-lying $3/2^+$ and $5/2^+$ states. A first search did not observe the $3/2^+$ [A. Spyrou \textit{et al.}, Phys. Rev. C 84, 044309 (2011)], however,  a resonance in $^{15}$Be was populated in a second attempt and determined to be unbound with respect to $^{14}$Be by 1.8(1) MeV with a tentative spin-parity assignment of $5/2^+$ [J. Snyder \textit{et al.}, Phys. Rev. C 88, 031303(R) (2013)]. 
\item[Purpose] Search for the predicted $^{15}$Be $3/2^+$ state in the three-neutron decay channel.
\item[Method] A two-proton removal reaction from a 55 MeV/u $^{17}$C beam was used to populate neutron-unbound states in $^{15}$Be. The two-, three-, and four-body decay energies of the $^{12}$Be $+$ neutron(s) detected in coincidence were reconstructed using invariant mass spectroscopy. Monte Carlo simulations were performed to extract the resonance and decay properties from the observed spectra.
\item[Results] The low-energy regions of the decay energy spectra can be described with the first excited unbound state of $^{14}$Be ($E_x=1.54$ MeV, $E_r=0.28$ MeV). Including a state in $^{15}$Be that decays through the first excited $^{14}$Be state slightly improves the fit at higher energies though the cross section is small.
\item[Conclusions] A $^{15}$Be component is not needed to describe the data. If the $3/2^+$ state in $^{15}$Be is populated, the decay by three-neutron emission through $^{14}$Be is weak, $\leq11\%$ up to 4 MeV. In the best fit, $^{15}$Be is unbound with respect to $^{12}$Be by 1.4 MeV (unbound with respect to $^{14}$Be by 2.66 MeV) with a strength of $7\%$.
\end{description}
\end{abstract}

\maketitle

The study of neutron-unbound states in atomic nuclei has been instrumental in probing the neutron drip-line \cite{ReviewBaumann}. Improvements in experimental techniques continue to push the discovery of nuclei farther from the valley of stability. Recent measurements of nuclei unbound by two neutrons include the first observations of $^{10}$He \cite{10HeKorsheninnikov}, $^{13}$Li \cite{13LiAksyutina}, $^{16}$Be \cite{16BeSpyrou}, and evidence for $^{26}$O \cite{26OLunderberg}.

The beryllium isotopes have been a fertile ground to study neutron-unbound states because of the availability of beams and low level densities. The heaviest beryllium isotope with a known bound state is $^{14}$Be \cite{14BeBowman} (though $^{13}$Be is unbound) and the heaviest isotope measured is the two-neutron unbound $^{16}$Be \cite{16BeSpyrou}.  Between these two isotopes lies $^{15}$Be, whose ground state still has yet to be confirmed. Shell model calculations have predicted low-lying $3/2^+$ and $5/2^+$ states located near each other \cite{15BePoppelier, 15BeSpyrou}. The first attempt to experimentally observe $^{15}$Be was performed using a two-proton removal from a $^{17}$C beam \cite{15BeSpyrou}. The ground state of $^{17}$C has been shown to have a $3/2^+$ spin and parity \cite{17CMaddalena} and the reaction used was expected to remove two $p$-shell protons while leaving the neutrons in their initial configuration. Very few $^{14}$Be fragments were observed in coincidence with the detected neutrons. Therefore it was concluded that the ground state of $^{15}$Be is likely to be located at a higher energy than the first unbound excited state in $^{14}$Be. This state at $E_r=0.28$ MeV would serve as an intermediate state for decay to $^{12}$Be. A second experiment was performed using a neutron stripping reaction on a CD$_2$ target with a $^{14}$Be beam to populate $^{15}$Be \cite{15BeSnyder}. A $^{15}$Be resonance was observed with a one-neutron decay energy of 1.8(1) MeV, a width of 575(200) keV and $\ell=2$. The result of this experiment alone does not answer the question of the location and the properties of the $^{15}$Be ground state, however, since the $3/2^+$ state could be above the first excited state in $^{14}$Be and still be lower than the observed resonance in $^{15}$Be. 

\begin{figure}
 \includegraphics[width=3.25in]{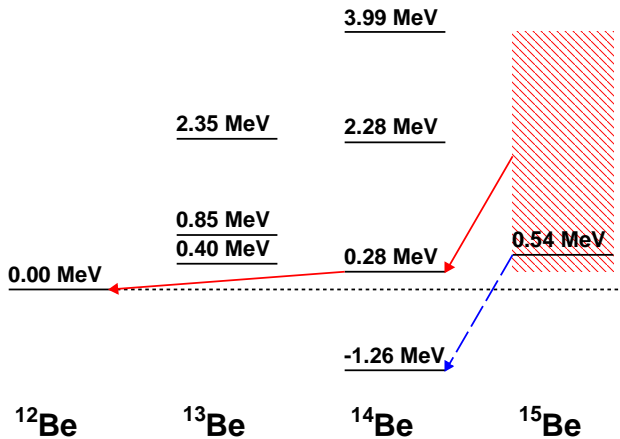}
 \caption{\label{fig:BeLevels} (Color online) Select low-lying levels in beryllium isotopes from \cite{13BeRandisi, 14BeAksyutina, 15BeSnyder}. The dashed blue arrow represents the one-neutron decay from $^{15}$Be to $^{14}$Be \cite{15BeSnyder} and the solid red arrows represent the suggested decay path in \cite{15BeSpyrou} searched for in this work. The red-shaded box represents the range of energies that were simulated.}
 \end{figure}

\begin{figure}
 \includegraphics[width=3.5in]{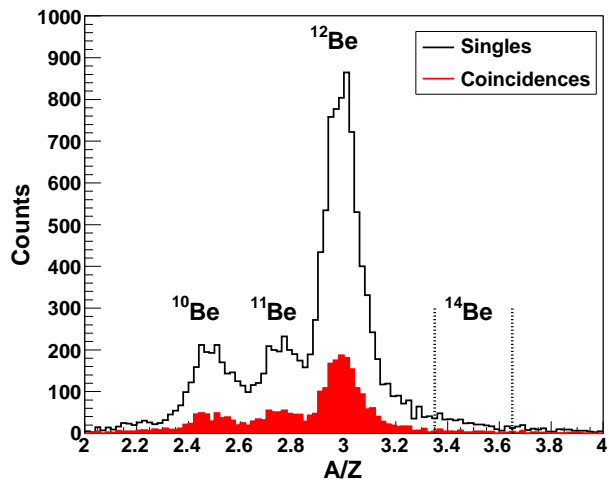}
 \caption{\label{fig:BRho} (Color online) The detected fragments deflected by the Sweeper magnet are shown by their mass-to-charge ratio. The black histogram represents fragments detected in singles mode and the filled red histogram shows fragments detected with a neutron coincidence. Two vertical dashed lines show where $^{14}$Be fragments should appear. This figure is adapted from Ref. \cite{15BeSpyrou}.}
 \end{figure}

Shell model calculations predict a large spectroscopic overlap (1.27 for $\ell=2$) between the predicted $3/2^+$ state and the first excited $2^+$ state in $^{14}$Be \cite{15BeSpyrou}. This calculation, paired with the non-observation of a low-lying state in the two-body decay of $^{15}$Be \cite{15BeSpyrou}, support the idea that the $3/2^+$ state could be observed in the three-neutron decay channel after decaying through the first neutron-unbound $2^+$ state in $^{14}$Be. This is shown schematically in Fig. \ref{fig:BeLevels} by the solid red lines. In the original analysis \cite{15BeSpyrou}, few $^{14}$Be fragments were observed while a large number of $^{12}$Be fragments were detected, as shown in Fig. \ref{fig:BRho}. In the present work, we analyze the possible $3n+^{12}$Be channel from the $2p$ removal experiment in Ref. \cite{15BeSpyrou} by reconstructing decay energies with multiple neutrons in coincidence with those $^{12}$Be fragments.

The experiment was performed at the National Superconducting Cyclotron Laboratory at Michigan State University. A 55 MeV/u $^{17}$C beam was produced from a $^{22}$Ne primary beam. This beam was focused onto a beryllium target where neutron-unbound states were populated. The emitted neutrons were detected with the Modular Neutron Array (MoNA) \cite{MoNA_NIM}. The fragments were deflected by the Sweeper dipole magnet \cite{sweeper} into charged-particle detectors. The measured four-momenta of the neutron(s) and fragment allowed the reconstruction of the decay energies by invariant mass spectroscopy. More experimental details can be found in Ref. \cite{15BeSpyrou}. 

\begin{figure*}
 \includegraphics[width=7.25in]{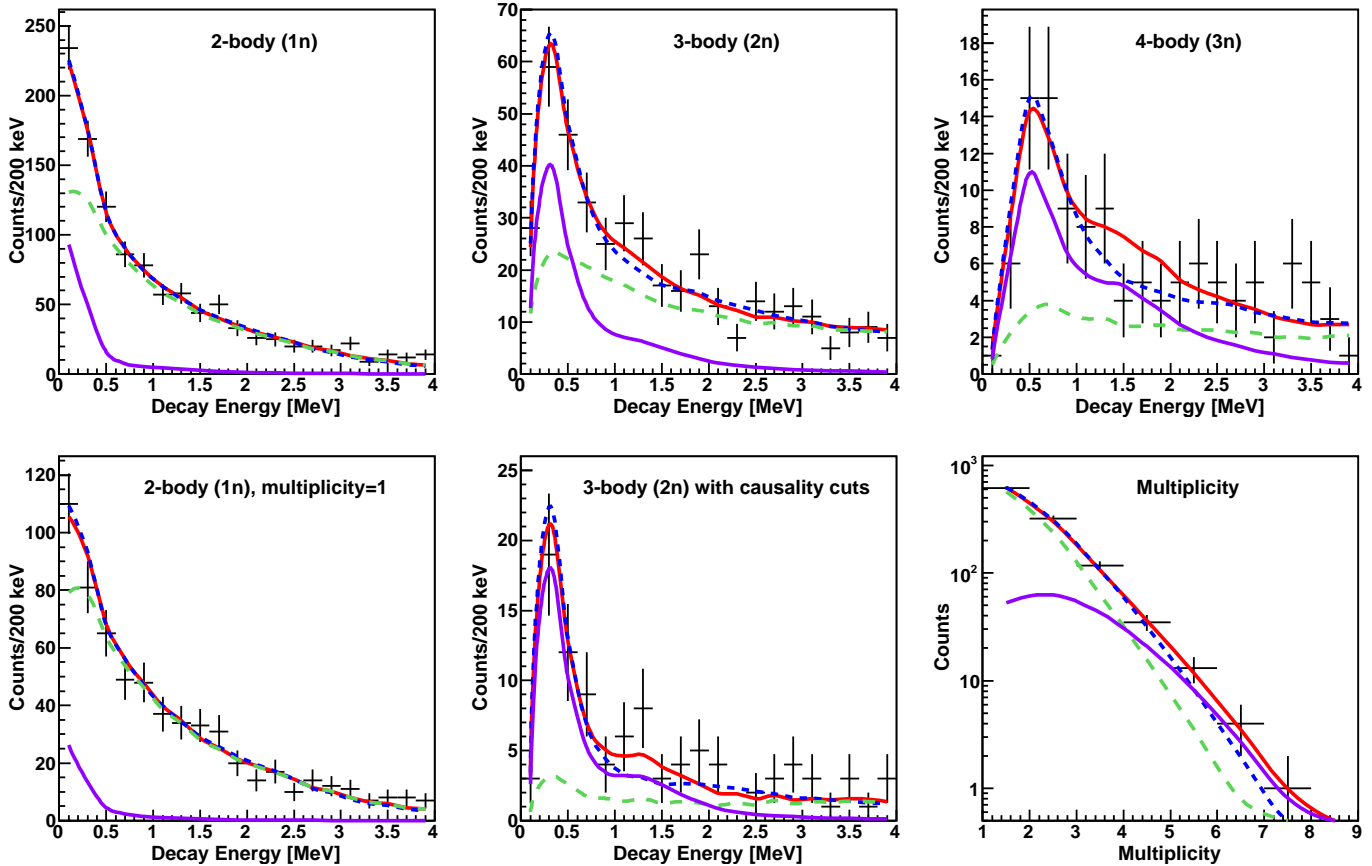}
 \caption{\label{fig:Spectra} (Color online) From left to right in the top row are the two-, three-, and four-body decay energies. From left to right in the bottom row are the multiplicity equal to one two-body decay energy, the three-body decay energy with causality cuts, and the multiplicity. The black crosses represent the experimental data. The dashed blue line represents the best fit if no $^{15}$Be is included. In this fit the 0.28-MeV state in $^{14}$Be is directly populated. The best fit is shown by the solid red line, which is the sum of the solid purple, $^{15}$Be at 1.4 MeV decaying through the $^{14}$Be 0.28-MeV state and dashed green lines made up of the higher lying $^{14}$Be states. }
 \end{figure*}

The multiplicity, or number of scintillator bars with a valid signal in a given event, is a useful parameter in searching for a state decaying by multiple neutron emission. In scenarios where unbound states decay by one neutron, the multiplicity distribution  is most likely to peak at one and decrease with each higher multiplicity value. This distribution is a result of scattering to multiple bars. When resonances decay by more than one neutron there is a greater probability of detecting multiple neutrons. This shifts the distribution toward higher multiplicities. To search for decays of $^{15}$Be to $^{12}$Be$+3n$, decay energies were reconstructed using one, two, and three neutrons with a coincident $^{12}$Be fragment in a time-ordered manner. In addition, restrictions were applied to the spectra to enhance or discriminate certain features. Applying a multiplicity-one restriction to a two-body decay energy spectrum enhances the signal from single neutron decays. To reduce cross-talk from multiple scattering between scintillator bars in MoNA, causality requirements were applied to the three-body decay energy spectrum  \cite{24OHoffman, 16BeSpyrou, 26OLunderberg, 13LiKohley, 12BeSmith}. The causality requirements used here restrict the first two hits to have velocities greater than 10 cm/ns and the distance between hits to be greater than 30 cm. While many valid two-neutron events are removed, the ratio of true two-neutron events to multiple scattering events is greatly improved. There were insufficient statistics to apply any causality cut to the four-body spectrum, thus it is dominated by cross-talk from one and two neutrons scattering in MoNA. Six spectra constructed from experimental data are shown in Fig. \ref{fig:Spectra}: the two-, three-, and four-body decay energy spectra, the multiplicity-one gated two-body spectra, the three-body spectra with causality cuts, and the multiplicity distribution. 

Monte Carlo simulations, which take into account experimental acceptances, resolutions, and efficiencies were performed to provide an interpretation of the data \cite{stmona_NIM}. To describe the data presented in this work, previously observed states in $^{13,14}$Be \cite{13BeRandisi, 14BeAksyutina} were included in the simulations. These states are shown in Fig. \ref{fig:BeLevels}. The energies and widths were fixed, only the relative strengths were free parameters. In addition to the simulation of these known states, the simulation of a new state in $^{15}$Be with freely varying energy, width, and relative strengths was included. The most sensitive parameter for this state was the decay energy which was varied from 50 keV to 4 MeV.

The simulated data are shown on top of the experimental data in Fig. \ref{fig:Spectra}. The experimental decay energy spectra feature a low-energy peak. The energy and width of this peak, taking into account the experimental resolution, can be described by the first excited neutron-unbound $2^+$ state in $^{14}$Be. This state is unbound with respect to $^{12}$Be by 0.28 MeV \cite{14BeSugimoto}. The higher-energy components of the spectra can be reproduced by higher-lying states reported in Ref. \cite{14BeAksyutina} (though, these parameters are not unique solutions to the fit). The 0.28-MeV state accounts for $\approx 11 \%$ of the total strength up to 4 MeV in the best fit when populated directly (no $^{15}$Be component). The sum of the 0.28-, 2.28-, and 3.99-MeV states is shown by the dashed blue line in Fig. \ref{fig:Spectra}.

\begin{figure}
 \includegraphics[width=3.25in]{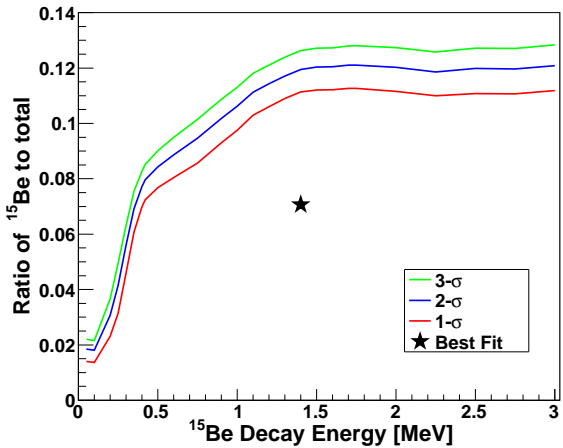}
 \caption{\label{fig:2DChi2} (Color online) The one- (red), two-(blue), and three-$\sigma$ (green) limits as a function of the ratio of $^{15}$Be to the total counts and the $^{15}$Be decay energy above $^{12}$Be. The best fit is indicated by the star.}
\end{figure}

A fit was also performed including a state in $^{15}$Be as indicated by the solid red line in Fig. \ref{fig:Spectra}. This fit is made up of the $^{15}$Be state decaying through the 0.28-MeV $^{14}$Be state (solid purple line) and higher-lying states in $^{14}$Be (dashed green line). The $\chi^2$ was calculated by comparing the simulated data to the experimental data in all six spectra. Using the $\chi^2$ value, the best fit (denoted by the star) and the one-, two-, and three-$\sigma$ limits as functions of the strength of $^{15}$Be and the $^{15}$Be decay energy are shown in Fig. \ref{fig:2DChi2}. To obtain ratios below the 0.28-MeV state in $^{14}$Be, four-body decays without intermediate states were simulated. The strength of this state for the best fit is $\approx 7 \%$ and the $1\sigma$ limit is at $\approx 11\%$.  The fit also includes 1.1$\%$ direct population of the $^{14}$Be state at 0.28 MeV. The simulations are insensitive to the width of the included $^{15}$Be state. For the fit shown, an $\ell=2$ resonance with a total width of 500 keV (arbitrarily chosen due to insensitivity) was used. As the energy of the state is increased, the fit approaches that of the 0.28-MeV state in $^{14}$Be being directly populated instead of acting as an intermediate state in the decay process. 

\begin{figure*}
 \includegraphics[width=7.25in]{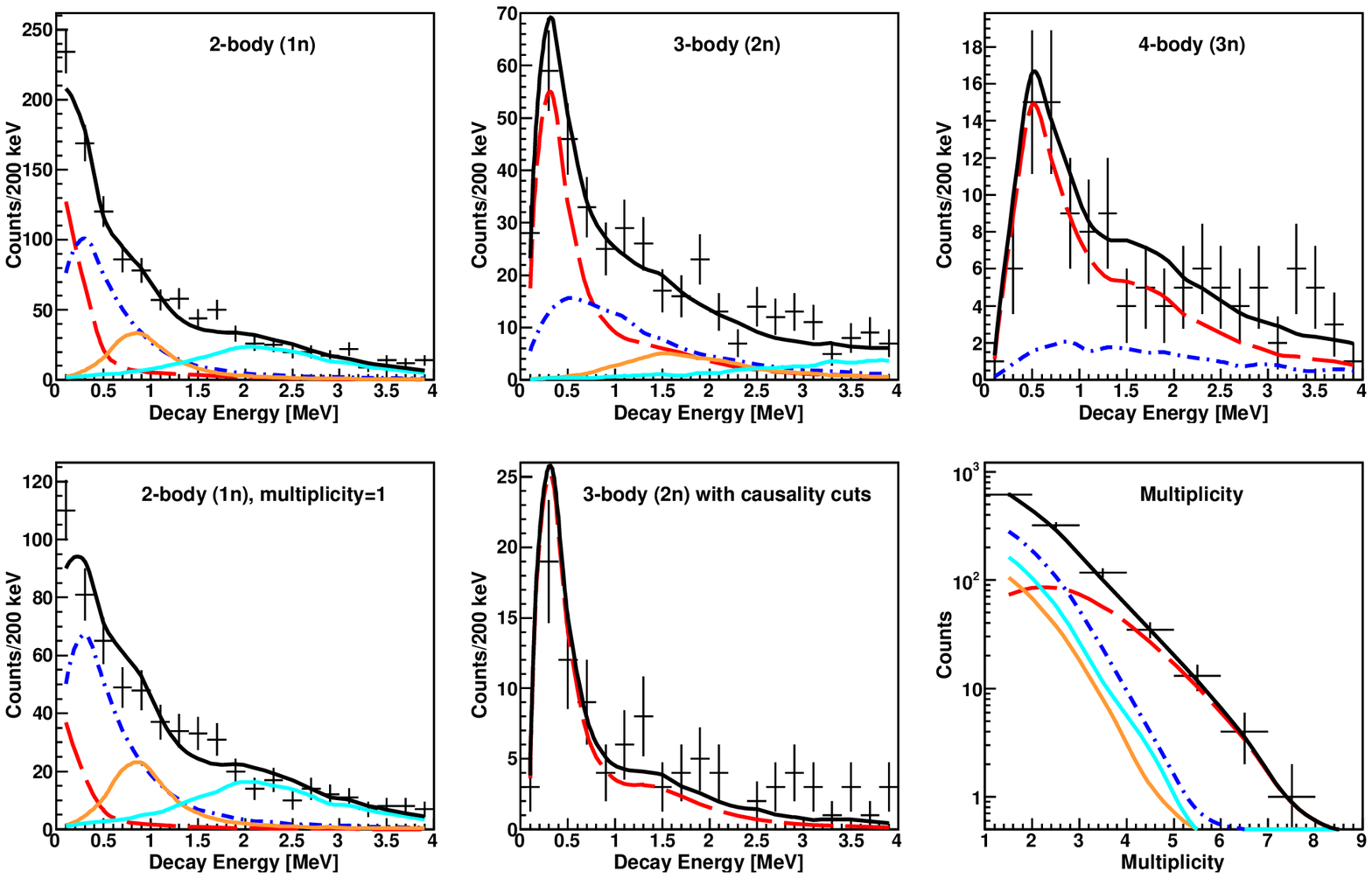}
 \caption{\label{fig:Spectra_alpha} (Color online) The six spectra are the same as Fig. \ref{fig:Spectra}, however, with a different fit. The solid black line represents the best fit made by summing the components of a state populated by two-proton knockout to $^{15}$Be at 1.6 MeV decaying through the 0.28-MeV state in $^{14}$Be (red dashed line) and $\alpha$ removal to $^{13}$Be at 0.40 MeV (dotted-dashed blue line), 0.85 MeV (solid orange line), and 2.35 MeV (solid light blue line).}
 \end{figure*}
 
 The overall cross section for populating $^{12}$Be was calculated from the singles data (events not requiring a coincident neutron hit), shown in Fig. \ref{fig:BRho}, to be $4.6\pm 1.1$ mb. The ratio of $^{15}$Be to the total extracted from the simulated decay energy spectra then implies that the two-proton removal cross section decaying to $^{12}$Be is smaller than 0.5 mb. This upper limit is a factor of two smaller than the calculated two-proton removal from $^{17}$C of 0.99 mb \cite{15BeSpyrou}. In addition, it would also suggest that the cross section for populating $^{14}$Be directly in a $2p1n$ removal reaction would be $\sim$ 4 mb, which would be larger than the two-proton removal cross section. Calculations predict the cross section to be an order of magnitude lower for the $2p1n$ removal process compared to the $2p$ removal \cite{PCTostevin}.
 
To resolve this discrepancy, a different process contributing to the overall cross section was considered. Recently, Sharov \textit{et al.} suggested that direct $\alpha$ removal might contribute to the population of $^{10}$He in the reaction of $^{14}$Be on $^9$Be at 59 MeV/u \cite{10HeSharov}. The corresponding process in the present reaction would lead directly to $^{13}$Be which then would populate $^{12}$Be by the emission of a single neutron. Similar evidence for such a process has also been observed in the analysis of the break-up data of $^{14}$Be populating $^{10}$He \cite{12HeJones}.
 
In order to test for this possibility, simulations were performed  where two reaction mechanisms were included: two-proton removal to $^{15}$Be and $\alpha$ removal to $^{13}$Be. Experimentally, $2p2n$ and $\alpha$ removals were indistinguishable. However, the cross section for $2p2n$ removal is expected to be even smaller than the $2p$ removal. For the two-proton removal, a single resonance in $^{15}$Be was assumed to decay via the 280-keV state in $^{14}$Be and the $\alpha$ removal was allowed to populate any of the three previously measured resonances in $^{13}$Be \cite{13BeRandisi}, shown in Fig. \ref{fig:BeLevels}. Peaks at similar energies have been measured in other previous works such as Refs. \cite{13BeSimon, 13BeKondo, 13BeAksyutina}. Free parameters were the energy and width of the state in $^{15}$Be, as well as all the relative strengths. The resulting best fit is shown in Fig. \ref{fig:Spectra_alpha}. In this fit, the energy of the $^{15}$Be state is 1.6 MeV. Based on the strengths of the states, the cross section for $\alpha$ removal would then be $\approx 4$ mb. Such a large cross section favors a direct $\alpha$ removal over a two-proton removal indicating a small two-proton removal cross section.
 
This work searched for the unobserved 3/2$^+$ state in $^{15}$Be. Based on comparisons of simulated data to experimental data, a $^{15}$Be state is not required. A fit with $2p1n$ removal to directly populate unbound states in $^{14}$Be is able to reproduce the decay energy and multiplicity spectra. If a state in $^{15}$Be is populated, the state is unbound to $^{12}$Be by 1.4 MeV with a strength $\leq11\%$ determined by the minimum $\chi^2$. Based on cross section calculations from the data for this scenario, the $2p$ removal cross section would be nearly an order of magnitude less than the $2p1n$ removal cross section. This disagrees with theoretical calculations which predict the $2p1n$ removal cross section to be an order of magnitude less than the $2p$ cross section. An alternative approach using simulated $2p$ removal to $^{15}$Be which decay by three neutrons and $\alpha$ removal to $^{13}$Be which decay by one neutron to $^{12}$Be was also able to describe the data. In this solution the $\alpha$ removal is the dominant reaction and again the cross section for the population of $^{15}$Be is small.

This work was supported by the National Science Foundation under Grants No. PHY06-07007, No. PHY09-69058, No. PHY11-02511, and PHY 12-05357. This material is also based upon work supported by the Department of Energy National Nuclear Security Administration under Award No. DE-NA0000979.


\bibliography{15Be}

\end{document}